\documentclass[a4paper,fleqn,usenatbib]{mnras}

\usepackage[T1]{fontenc}
\usepackage{times,txfonts} 

\usepackage{graphicx}	
\usepackage{natbib}
\usepackage{multicol}        
\usepackage{hyperref}

\title[GeMS+GMOS imaging performance]{FIRST PERFORMANCE OF THE GeMS + GMOS SYSTEM. Part 1. Imaging.\footnote{Based on observations obtained at the Gemini Observatory, which is operated by the Association of Universities for Research in Astronomy, Inc., under a cooperative agreement
with the NSF on behalf of the Gemini partnership: the National Science Foundation (United
States), the National Research Council (Canada), CONICYT (Chile), the Australian Research 
Council (Australia), Ministério da Ciência, Tecnologia e Inovação (Brazil) and
Ministerio de Ciencia, Tecnología e Innovación Productiva (Argentina).}}

\author[P. Hibon et al.]{
Pascale Hibon,$^{1,8}$
Vincent Garrel,$^{1}$
Benoit Neichel,$^{1,4}$
Benjamin Prout,$^{2}$
Francois Rigaut,$^{2}$
\newauthor Alice Koning,$^{3}$
Eleazar R. Carrasco,$^{1}$
German Gimeno,$^{1}$
Peter Pessev$^{5,6,7}$
\\
$^{1}$Gemini Observatory, Colina El Pino S/N, La Serena, Chile\\
$^{8}$ European Southern Observatory, Alonso de Cordova 3107, Casilla 19001, Santiago, Chile\\
$^{2}$Australian National University, Barry Dr Acton Australian Capital Territory 0200, Australia\\
$^{3}$University of Victoria, 3800 Finnerty Rd Victoria, BC V8P 5C2, Canada\\
$^{4}$Aix Marseille Universit\'e, CNRS, LAM (Laboratoire d'Astrophysique de Marseille) UMR 7326, 13388, Marseille, France\\
$^{5}$Instituto de Astrof\'isica de Canarias,  C/ Via Lactea, s/n, 38205, La Laguna, Tenerife, Spain\\
{$^{6}$Departamento de Astrof\'isica, Universidad de La Laguna, 38206 La Laguna, Tenerife, Spain}\\
$^{7}$Gran Telescopio Canarias (GRANTECAN), 38205 San Crist\'obal de La Laguna, Tenerife, Spain\\
}

\date{Accepted March 23rd 2016. Received YYY; in original form ZZZ}

\pubyear{2016}

\begin{document}
\label{firstpage}
\pagerange{\pageref{firstpage}--\pageref{lastpage}}
\maketitle

\begin{abstract} 
During the commissioning of the Gemini MCAO System (GeMS), we had the opportunity to obtain data with the Gemini Multi-Object Spectrograph (GMOS), the most utilised instrument at Gemini South Observatory, in March and May 2012. Several globular clusters were observed in imaging mode that allowed us to study the performance of this new and untested combination. GMOS is a visible instrument, hence pushing MCAO toward the visible. We report here on the results with the GMOS instruments, derive photometric performance in term of Full Width Half Maximum (FWHM) and throughput. In most of the cases, we obtained an improvement factor of at least 2 against the natural seeing.  This result also depends on the Natural Guide Star constellation selected for the observations and we then study the impact of the guide star selection on the FWHM performance. We also derive a first astrometric analysis showing that the GeMS+GMOS system provide an absolute astrometric precision better than 8mas and a relative astrometric precision lower than 50 mas. 
\end{abstract}


\section{Introduction}
\label{intro}
Data from ground based telescopes suffer from the effects of turbulence in the atmosphere. The first systems for sensing and correcting these atmospheric aberrations, known as adaptive optic (AO) system, were proposed in 1953 by \cite{Babcock}. Adaptive optics correct incoming light from distant celestial bodies, typically very dim, by using a relatively bright natural guide star as a reference. With AO, the resolution of the images improves dramatically as long as the targets are close enough to a bright reference star. However, sufficiently bright stars are not available in all parts of the sky and typically less than 10\% of the sky can benefit from AO corrections. To extend the use of AO systems to the whole sky, artificial stars also called laser guide stars (LGS) have been developed (\cite{Foy}). 
Using laser guide stars does not come without its own drawbacks: due to the finite altitude of the LGS, some turbulence goes unseen by the WFS, limiting performance of the AO system. Using multiple-LGS not only allows to solve this so-called cone effect, but can also be used to increase the size of the corrected FoV.
Because of back and forth laser propagation, Tip Tilt modes stay undetermined, a significant detrimental effect for AO correction as Tip Tilt are the mode containing the most energy from atmospheric turbulence. (optional example phrase:) For example, in K-band and good seeing conditions (r0=60cm in K), the image motion (TT only) is roughly 0.2" rms, i.e 3.6 times the size of the perfect Airy Disk PSF. Rigaut and Gendron (1992) proposed to correct this issue by adding a second wavefront sensor guiding on natural guide star. This additional WFS can be as simple as a quadcell, allowing to guide on very faint guide stars. In this scheme, sky coverage is almost complete as the presence of such faint stars is quite common even around galactic poles. 
Different AO methods have been developed to allow observations over a wide field of view : Ground Layer Adaptive Optics (GLAO) \citep{Rigaut2002, Tokovinin2004}, Multi-Object Adaptive Optics (MOAO) \citep{Hammer2004} and Multi-Conjugate Adaptive Optics (MCAO)\citep{Beckers1988, Rigaut2001}. In GLAO, only the atmospheric turbulence close to the ground is corrected. This correction enhances the resolution over a wide field, as the light from every object in the sky pass through the same low layer of turbulence before reaching the telescope. However, the correction provided by a GLAO system is only partial, and usually does not reach the diffraction limit of the telescope. In order to further improve the performance over the full field, one need to add corrective elements (i.e. deformable mirrors) to compensate for the high altitudes turbulent layers. 
MCAO has first been demonstrated with the Multi-Conjugate Adaptive Optics Demonstrator (MAD) \citep{Marchetti2003}, and now used in regular operation at Gemini-South, with the GeMS instrument \citep{Rigaut2014, Neichel2014}.

GeMS is the dedicated AO facility at the Gemini South Telescope located in Cerro Pach\'on. It is the first instrument using a fixed 5 laser sodium guide stars (LGS) asterism, in addition of 3 natural guide stars (NGS), to compensate the optical distortions induced by atmospheric turbulence over a 2' wide field of view  \citep{Rigaut2014}. The compensation can be achieved using two deformable mirrors conjugated at the ground layer and at 9km altitude. Over the full field, it can provide a uniform close to diffraction-limit Point Spread Function (PSF) in the near infrared bands (J to K), 5 to 10 times larger compared more classic Single Conjugated Adaptive Optic Systems (SCAO) or Laser Conjugated Adaptive Optic Systems (LGSAO) \citep{Neichel2014}. GeMS has been routinely in operation since mid-2013 in combination with the near-infrared Gemini South Adaptive Optics Imager (GSAOI) \citep{McGregor2004}. \\

At the present time, compensating enough the atmospheric turbulence in order to reach the diffraction limit in the visible bands still remains a challenging endeavor. This can now be accomplished over a very small field of view (10" maximum) around bright stars with the upcoming generation of Extreme AO system \citep{Macintosh2006,Dohlen2006,Esposito2010,Jovanovic2015,Dekany2006,Close}. The sky coverage for these instrument is logically extremely small. Over a wide field of view, different approaches have been taken such as reaching the diffraction-limit with a relatively small telescope and some post-processing (1.5m RoboAO,\cite{Baranec}, LAMP \cite{Law2009}) or such as delivering a partial correction with larger telescopes. The Southern Astrophysical Research Telescope (SOAR) adaptive module (SAM) has been the first one in this category. Correcting for the ground layer turbulence only and using a UV Rayleigh LGS guide star and 2 NGS for tip tilt, it can deliver a PSF whose FWHM is down to 280mas in i-band over a square field of view of 3 arcmin \citep{Tokovinin}.\\
During the commissioning of GeMS, concurrently to GSAOI, we also used GMOS \citep{Hook} in its imaging mode to explore the performance of this MCAO system in the visible bands. We report with this publication the first performance in term of FWHM estimation, photometric and astrometric accuracies. In the first part, we are presenting the observations and the data reduction method. In a second part, we are explaining the FWHM performance for the GeMS+GMOS datasets. Then we estimated the throughput and the zero-point magnitudes obtained. We also examine the astrometric performance reached for these observations. Finally, we are discussing the possible science applications for such a unique system.

\section{Observations and Data Reduction}

\subsection{GMOS: Gemini Multi-Object Spectrograph}
GMOS is a spectro-imager working in the visible bands. It remains the most requested and used instrument at Gemini South Observatory with 72\% of the requested observing time in 2012.
In its spectrograph mode, long-slit, multi-slit and one Integral Field Unit (IFU) are available. The imaging mode covers a 5.5' x 5.5' field of view (FoV) over three CCD chips with a pixel scale of 79 mas. The three CCD chips form a 6144 x 4608 pixel array, with two gaps of about 37 pixels separating the detectors \citep{Hook}. In its imaging mode, GMOS-S has six standard broad band filters: u-band (336-385nm), g-band (398-552nm), r-band (562-698nm), i-band (706-850nm), CaT-band (780-933nm) and z-band ($\geq$848nm). The present paper focus on GMOS-S's imaging capabilities.
\subsection{GeMS: Gemini Multi-Conjugate Adaptive Optics System}

The Gemini Multi-Conjugate Adaptive Optics system, a.k.a. GeMS, is the first multi-laser guide stars system offered to the astronomical community \citep{Rigaut2014, Neichel2014}.
GeMS introduces 3 main optical changes:
\begin{itemize}
	\item the throughput: the losses due to reflections have been evaluated to about 30
	\item the f-ratio: GeMS modifies the native telescope f-ratio of f/16 to f/33.2 
	\item the field of view: GeMS field of view is a disk of 2 arcmin diameter.
	\item the current beamsplitter cuts the visible light at 750 to 800nm.
\end{itemize}

\subsection{Observations}
During the commissioning of GeMS, we had the opportunity to obtain data with GMOS in March and May 2012. GeMS change the native f/ratio of the telescope, from an f/16 beam to an f/33.2, hence, the imaging FoV of GMOS through GeMS is reduced to 2.5'x2.5', and with the pixel scale also reduced to 35.9mas.
Also, because GeMS as been designed to work in the NIR (see Section~\ref{results}), only the reddest filter of GMOS can be used. These are the i-band (706-850nm partially), CaT-band (780-933nm) and z-band ($\geq$848nm).  Table~\ref{tab:1} summarised the differences when using GMOS and the system GeMS+GMOS.\\

\begin{table}
\caption{Table comparing the specificities of GMOS and GeMS+GMOS.}
\label{tab:1}       
\begin{center}
\begin{tabular}{llll}
\hline
System & Field of View & Pixel Scale & Available Broad Filters  \\
\hline 
GMOS & 5.5' x 5.5' & 73 mas & u, g, r, i, CaT, z\\
GeMS+GMOS & 2.5' x 2.5' & 35.9 mas & i, CaT, z \\
\hline
\end{tabular}
\end{center}
\end{table}
We observed 15 targets including nine globular clusters with the GeMS+GMOS system during March and May 2012. We will now focus on the globular clusters. They were selected as best choice targets as they contain numerous bright stars that can be used for the NGS constellation. The quantity of stars in globular clusters is also a great advantage to help for studying the different performance (FWHM, astrometry for example) of the system and its variability over the field of view. Table~\ref{tab:gc} summarised the observed targets and the observation characteristics. We will not present the whole performance analysis for every globular cluster. Instead, we chose to show only the most relevant results.\\

\begin{table*}
\caption{Table summarising the observations obtained with GeMS+GMOS.}
\label{tab:gc}       
\begin{center}
\begin{tabular}{lllllll}
\hline
Targets & Dates & Filters & Exposure &  Offset & Binning & Average FWHM (mas)\\
NGC 2849 & 2012-03-15 & z & 27 x 5s  & Yes & 2 x 2 & 236.94 \\
NGC 3201 & 2012-03-13 & z & 12 x 60s & No & 2 x 2 &  466.7\\
NGC 3244 & 2012-03-16 & i & 31 x 5s & No & 2 x 2 &  165.14\\
NGC 4590 & 2012-03-15 & i, z, CaT & 56 x 5s & Yes & 2 x 2 & CaT:201.04, i:222.58, z:215.4 \\
                 & 2012-05-10 & i & 33 x 5s & No & 2 x 2 &  514.09\\
NGC 5139 & 2012-03-16 & i, z, CaT & 20 x 90s & Yes & 1 x 1 &  CaT:75.39, i:61.03, z:71.8\\
NGC 5286 & 2012-03-09 & z & 5 x 5s & No & 1 x 1 &  301.56\\
                 & 2012-05-10 & i & 20 x 5s & No & 2 x 2 &  502.6\\
NGC 5408 & 2012-03-10 & z & 4 x 10s & No & 2 x 2 &  122.06\\
NGC 6369 & 2012-03-14  & z, i & 9 x 120s & Yes & 1 x 1 &  i:183.09, z:197.45\\
                 &                      & i & 2 x 240s & Yes & 1 x 1 & 183.09 \\
NGC 6496 & 2012-05-09 & i & 25 x 5s & No & 2 x 2 & 567.22 \\
                 & 2012-05-10 & i & 6 x 5s & No & 2 x 2 &  552.86\\
                 & 2012-05-10  & i & 4 x 15s & No & 2 x 2 &  782.62\\
                 & 2012-05-10 & i & 24 x 15s & Yes & 2 x 2 &  574.4\\
                 & 2012-05-10 & z, CaT & 21 x 10s & Yes & 2 x 2 &  CaT:746.72, z:660.56\\
CENTAURUS & 2012-03-11 & z & 14 x 5s & No & 2 x 2 &  1191.88\\
                 & 2012-03-12 & i, z, CaT & 13 x 60s & No & 2 x 2 & CaT: 71.8, i: 66.77, z:72.87 \\
                 &                      & z & 26 x 10s & Yes & 1 x 1 &  195.29\\
                 & 2012-03-13 & z & 17 x 5s & No & 2 x 2 &  174.47\\
                 & 2012-03-14 & i, z & 6 x 120s & Yes & 1 x 1 & i: 171.84, z: 137.83 \\
CIRCINUS & 2012-03-12 & z & 12 x 15s & No & 2 x 2 &  466.7\\
                &                      & z & 7 x 45s & Yes & 2 x 2 &  287.2\\
                & 2012-03-13 & z & 21 x 5s & No & 2 x 2 &  172.32\\
                & 2012-03-14 & z & 6 x 5s & No & 1 x 1 &  323.1\\
                & 2012-03-16 & z & 4 x 5s & No & 2 x 2 &  192.42\\
                &                      & i & 4 x 600s & Yes & 2 x 2 &  177.34\\
IC 4296 & 2012-03-15 & i, z, CaT & 42 x 60s & Yes & 2 x 2 & CaT: 152.21, i: 137.85, z: 160.11  \\
M93 & 2012-03-11 & i, z& 5 x 5s & No & 2 x 2 & i: 538.5, z: 646.2\\
ORION & 2012-03-12 & i, z, CaT & 23 x 60s & Yes & 2 x 2 & CaT: 269.25, i: 308.74, z: 265.66 \\
            & 2012-03-15 & i & 14 x 5s & Yes & 1 x 1 &  251.3\\
SAGITTARIUS & 2012-03-11 & z & 6 x 30s & No & 2 x 2 & 193.86 \\
                     & 2012-03-16 & z & 19 x 300s & Yes & 2 x 2 &  172.32\\
\hline
\end{tabular}
\end{center}
\end{table*}


GeMS AO has not been designed for optical bands and therefore it will not reach the diffraction limit of the telescope under no less than exceptionally good seeing conditions. Performance in terms of PSFs FWHM are expected to vary both in absolute from excellent to quite poor but they also may lack of uniformity inside the GMOS field, a detrimental effect in most science cases. This lack of homogeneity is due non corrected anisoplanatism.

\begin{figure*}
\begin{center}
\includegraphics[width=9.5cm]{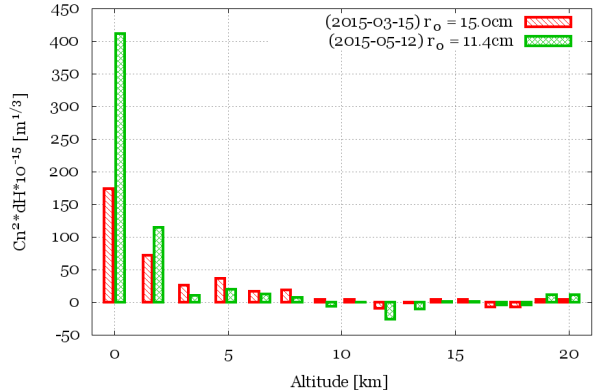}
\vspace{-1cm}
\caption{Cn2 turbulence profiles. The bars represent the turbulence profiles obtained in pseudo-open loop profile. The dashed bars (red) are from March 15th 2012 and the checked bars (green) from May 10th 2012. For visibility purpose, the dashed bars were shifted by -0.25km and the checked bars by +0.25km. The negative value we can remark at an altitude of 12km and above, can be either just noise or a bad estimation of the outer-scale. Generally is the latter:  the outer-scale is used to deconvolve the raw profile using the autocorrelation of the slopes, which is a function of the outer scales. We kept the negative values as it gives us information on how reliable our results are. } 
\label{fig:Cn2}
\end{center}
\end{figure*}

The wind profiling method, described in \cite{Cortes}, based on the estimation of the refractive index structure parameter $\mathrm{C}_{\mathrm{n}}^{2}$ allows us to assess the dynamical turbulence structure, and to explain then the different performance obtained for these datasets.
Figure~\ref{fig:Cn2}  presents the turbulence profiles obtained on sky on March 15th 2012 and May 10th 2012 for a pseudo open-loop (pseudo OL) measurements, represented in yellow. 
The seeing measured during these nights is $\mathrm{r}_{0}=11.4\mathrm{cm}$ (seeing$\sim$0.88'') and  $\mathrm{r}_{0}=14.8\mathrm{cm}$ (seeing$\sim$0.68''), respectively.
We remarked also the presence of a strong dome seeing component, which is part of the ground layer (altitude 0km) turbulence.
The better FWHM results obtained during March 15th 2012 than during the May 10th night (see Table~\ref{tab:gc}) can be explained by the difference of $\mathrm{r}_{0}$ values. However the presence of stronger turbulences at higher atmospheric layers affects more the homogeneity of the AO performance during the first night than during the latter one.\\
In order to improve GeMS + GMOS performance, as Gemini is a queue based telescope, we could use this information to know whether or not we are in conditions for observing with this system or not. Optimal conditions for GeMS+GMOS will be of course with the minimum high layer turbulence level, similar but slightly better than an equivalent GLAO system.
\subsection{Data Reduction}

\subsubsection{Data reduction method}
To process the GeMS+GMOS data, we use the Gemini IRAF package.\\
At the end of each night, bias frames were taken for every GMOS configuration.  Twilight flats were also obtained either at the beginning or at the end of the nights. We process the bias frames by overscan subtraction, trimming, and stacking to produce a master bias frame for each night. A very similar procedure was applied to the twilight frames to produce a master flat frame. The science frames are bias subtracted and flat field corrected in the standard way. \\
As we obtained data for the redder part of the GMOS bandpass, most of the frames were contaminated by fringes.
Fringe frames were obtained during March and May 2012. The fringe correction and the sky subtraction were realised simultaneously using \textit{girmfringe}/IRAF task. Finally, the cleaned individual frames are reconstructed as single extension images using the \textit{gmosaic}/IRAF task.\\
Some issues affect the stacking of these data.
\begin{itemize}
\item CANOPUS has a dynamic distortion which changes every night. It is therefore challenging to stack the datasets taken with offsets. This effect does not prevent us to stack the dataset taken without offsets.
\item Due to the exceptional GeMS+GMOS combination, the WCS is not presented in the data headers. A WCS calibration with GeMS+GMOS has not been obtained.
\end{itemize}

\subsubsection{Astrometric calibration}
A good World Coordinate System (WCS) calibration will depend on two main points : the availability of appropriate reference catalogs and the AO corrections.\\
The astrometric accuracy for the MCAO system and the instrument it feeds has already received much attention \citep{Trippe2010,Meyer2011,Schoeck2013,Neichel2014b,Lu2014} as the PSF uniformity over a large field and the ability to actively control the plate scale can significantly reduce the largest astrometric errors encountered in previous AO systems. However this requires that the systemic residual distortions are well under control: the system should be carefully calibrated in the world coordinates system and the systematic errors should be kept low. 
During the early commissioning, priority was given to explore the AO performance and test functionality of different MCAO subsystems rather than to carefully calibrate the astrometric performance. For example, the plate scale control was not yet applied  at this time. The targets used to assess performance were central part of globular clusters, i.e. very crowded fields, a perfect benchmark to check the PSF uniformity over the GeMS field of view. In this paper, we decided to focus on astrometric performance reached in single frames. We are lacking precise unconfused references stars to derive an accurate WCS as most of the catalogue can only offer a precision of about 1" \citep{Lasker}, i.e. about 27 times larger than the plate scale of GeMS+GMOS (0.0359 arcsec per pixel).
Using the Mikulski Archive for Space Telescopes (MAST), we found images from the Hubble Space Telescope (HST) corresponding to the globular clusters observed with GeMS+GMOS. We first identified the brightest stars found in both frames and run the \textit{ccmap}/IRAF task to compute the plate solution from a list of  matched pixel  and  celestial coordinate susing a polynomial function of order 2. Once this first WCS approximation was registered in the header of the GeMS+GMOS, we used the SIMBAD database \citep{Wenger2000} to identify as much objects as possible in the field. We then run the \textit{msctpeak}/IRAF task, from the \textit{mscred} package, with a funcion polynomial of order 4 with a sky projection combining the tangent plate projection and polynomials, called TNX WCS projection. After running this task twice on the Z-band images of NGC6496, we obtained an acceptable WCS solution with a WCS rms of 50mas (corresponding to 0.7-1 pixel approximately) in both direction. On the Z-band images of NGC3201, the best solution was achieved with a WCS rms of 90mas (corresponding to 2-3 pixels approximately). In the case of NGC 5139, the best achieved solution in both direction is 65-76mas in I-band, 80-95mas in CaT-band and 76-90mas in Z-band.



\section{FWHM Performance} 


\subsection{Measurement method}
The PSF describes the two-dimensional distribution of light
in the telescope focal plane for astronomical point sources. 
The PSF can be characterised by its FWHM or by the diameter enclosing a given percentage of the total brightness, thus describes the angular resolution achieved in an observation.\\
Images taken with AO should show a dramatic improvement gained through the use of adaptive optics and the FWHM is the first measure that show the difference with and without AO.\\ 
The star detection and the FWHM values are obtained by running SExtractor \citep{Bertin} on the reduced individual images. The SExtractor parameters were optimised to allow the detection of faint punctual objects. We measured the FWHM for every processed single exposures for every target observed with the GeMS+GMOS system.\\

\begin{figure*}
\resizebox{1.5\columnwidth}{!} {\includegraphics{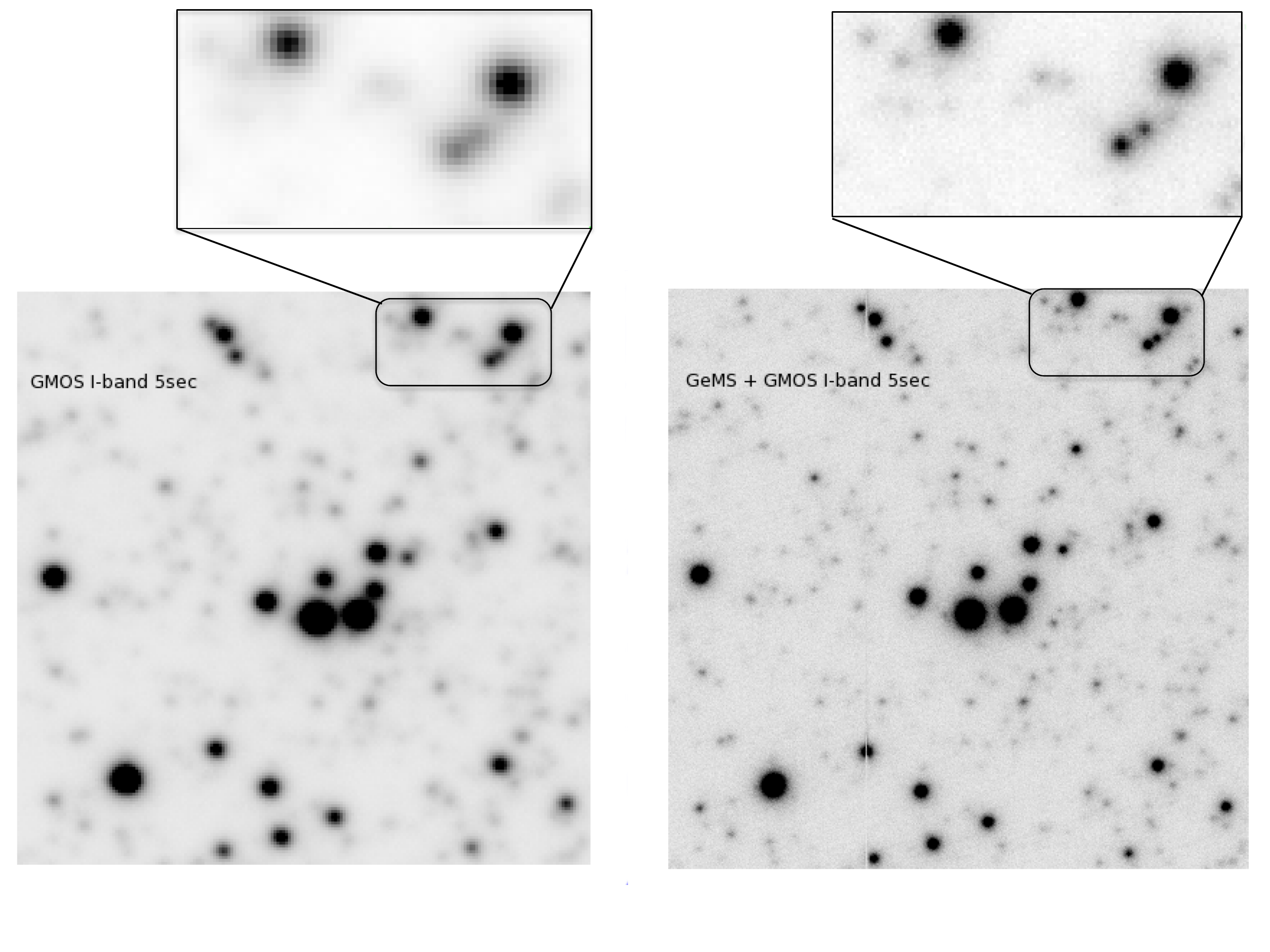} }
\vspace{-0.7cm}
\caption{Visual Comparison of the Image quality on a section of NGC4590 taken in I-band for 5sec exposure. Left : Taken with GMOS. Its average FWHM is 0.7arcsec. Right : Taken with GeMS+GMOS. Its average FWHM is 0.35arcsec}
\label{fig:illus}       
\end{figure*}
Figure~\ref{fig:illus} shows two single 5s exposure of a the globular cluster NGC4590 observed in the red, at I band ($\lambda$=780nm), one with GMOS (left) and one with GeMS+GMOS (right), during the same night. Faint and crowded sources can now be identified. GeMS provided a factor 2 of improvement in FWHM.

\subsubsection{FWHM maps}

\begin{figure*}
\resizebox{1.5\columnwidth}{!} {\includegraphics{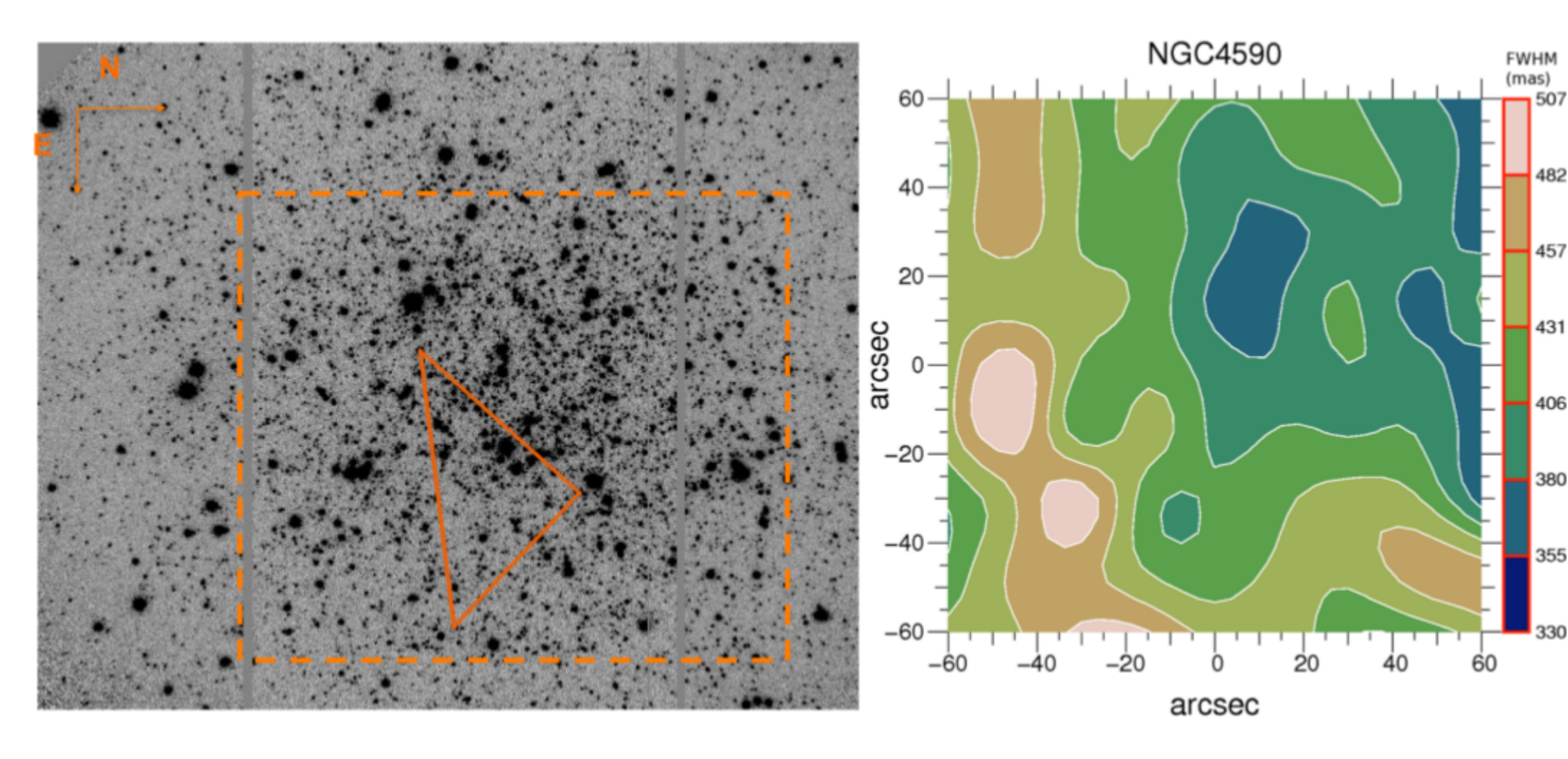} }
\caption{Left : Field image of NGC 4590 with the N-E orientation, \textbf{in I-band}. The NGS constellation and the studied area marked. Right : FWHM map.}
\label{fig:fwhm1}       
\end{figure*}

\begin{figure*}
\resizebox{1.5\columnwidth}{!}{\includegraphics{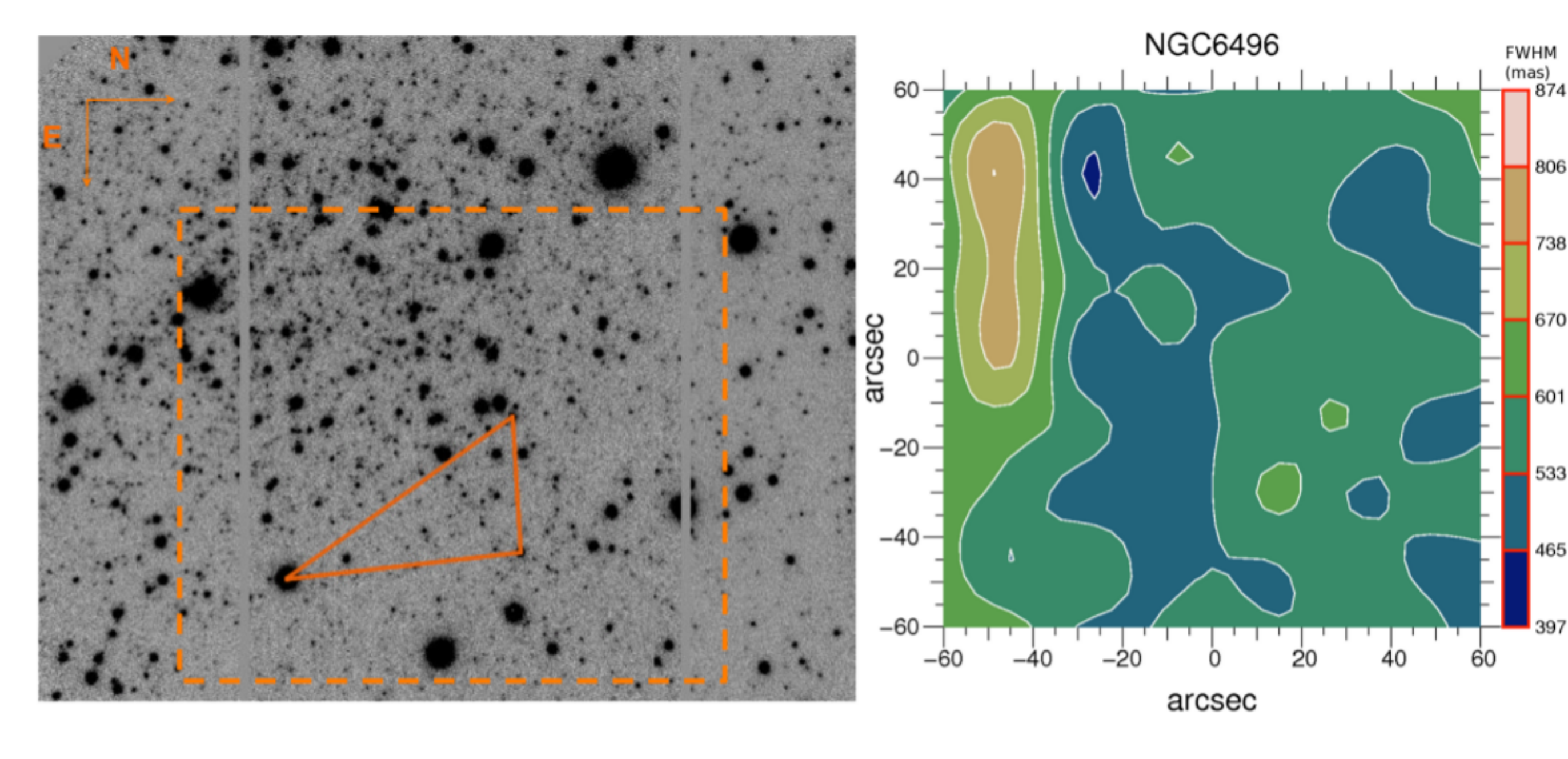} }
\caption{Left : Field image of NGC 6496 with the N-E orientation, \textbf{in I-band}. The NGS constellation and the studied area marked. Right : FWHM map.}
\label{fig:fwhm2}       
\end{figure*}

\begin{figure*}
\resizebox{1.5\columnwidth}{!}{\includegraphics{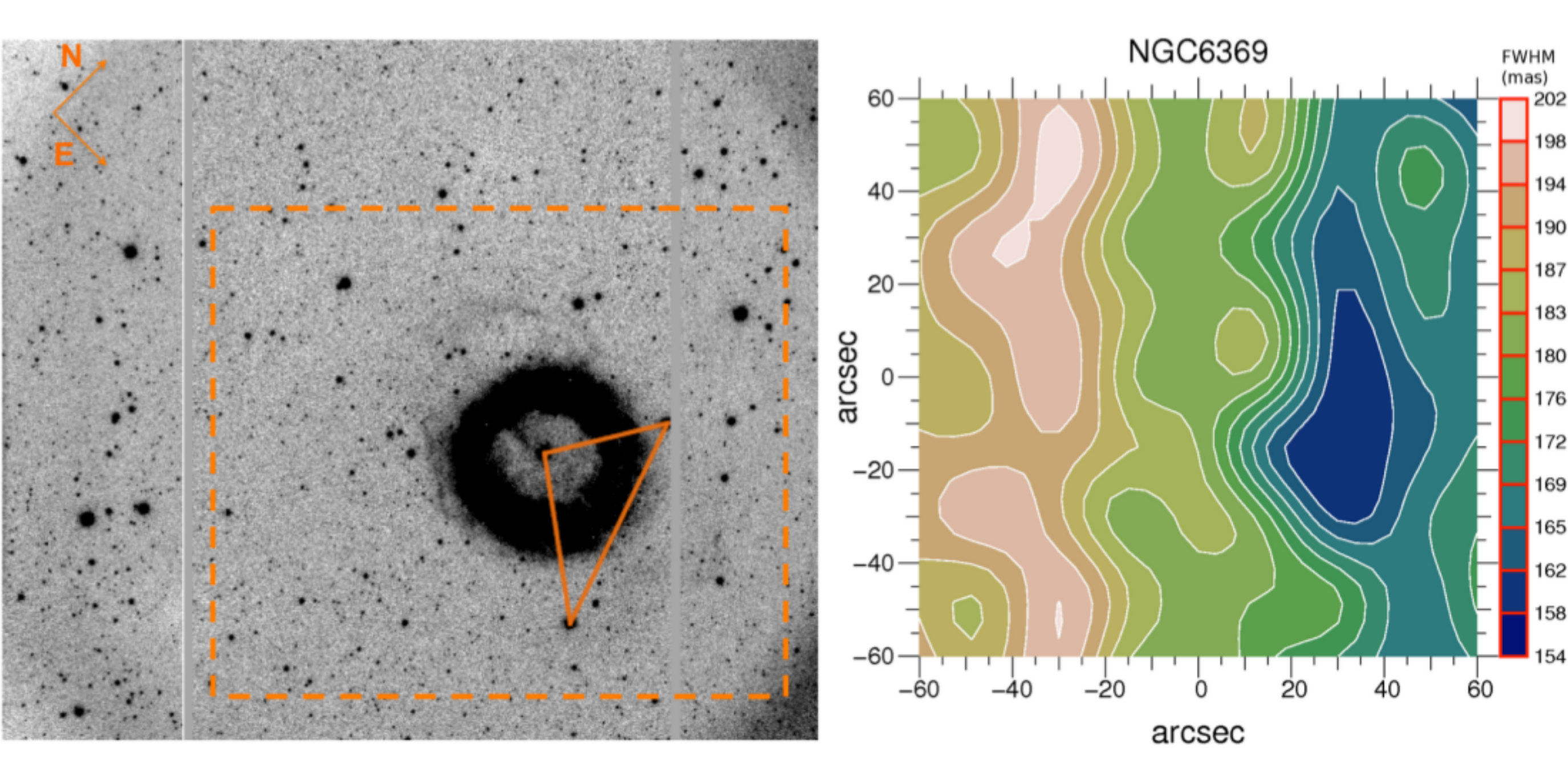} }
\caption{Left : Field image of NGC 6369 with the N-E orientation, \textbf{in Z-band}. The NGS constellation and the studied area marked. Right : FWHM map.}
\label{fig:fwhm3}       
\end{figure*}

\begin{figure*}
\resizebox{1.5\columnwidth}{!}{\includegraphics{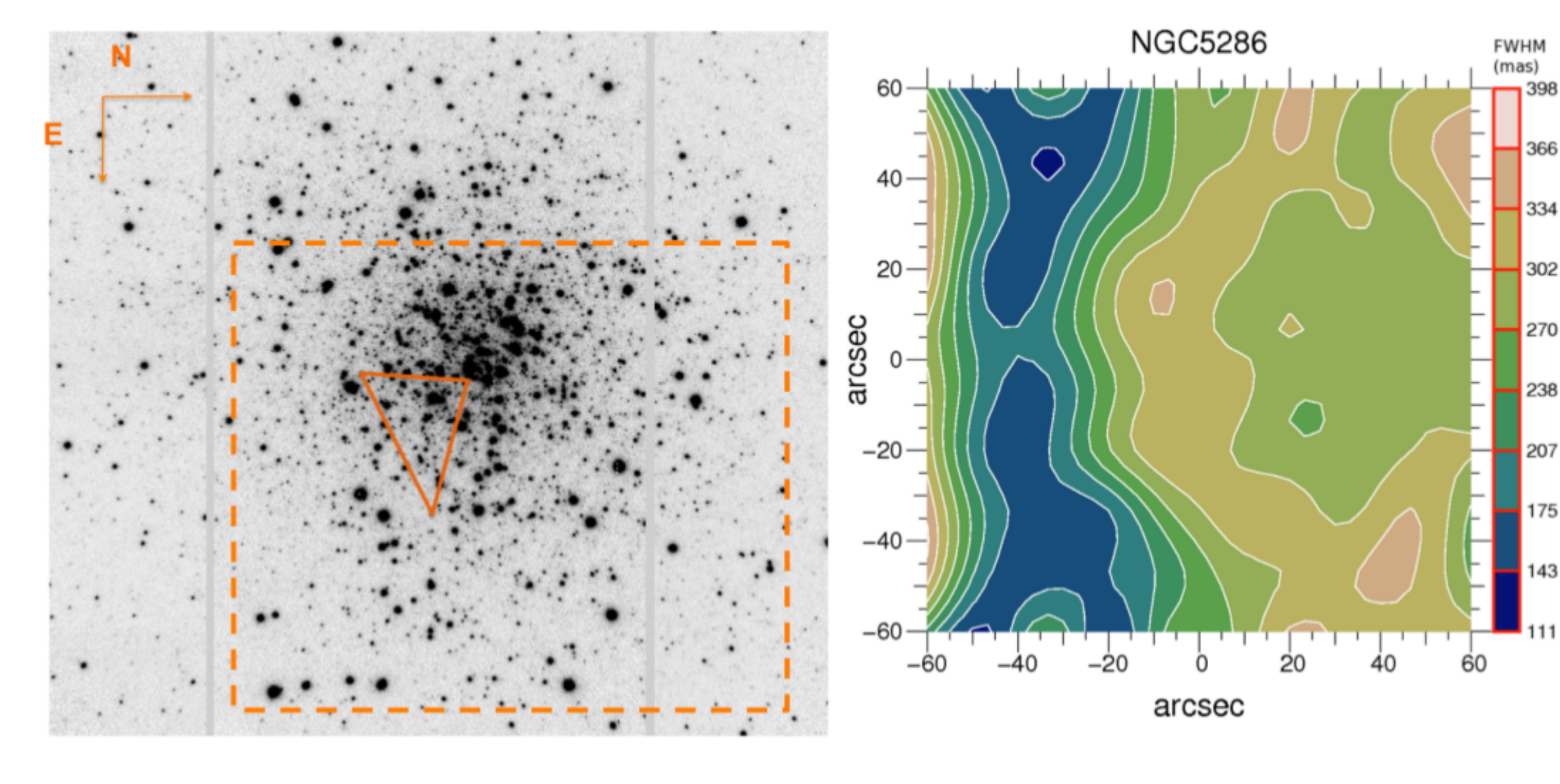} }
\caption{Left : Field image of NGC 5286 with the N-E orientation, \textbf{in Z-band}. The NGS constellation and the studied area marked. Right : FWHM map.}
\label{fig:fwhm4}       
\end{figure*}

Figures~\ref{fig:fwhm1}, ~\ref{fig:fwhm2}, ~\ref{fig:fwhm3}, and ~\ref{fig:fwhm4} present the field images with the Natural Guide Star (NGS) Constellation and the studied area, for which the FWHM has been determined, and the FWHM maps for four different globular clusters observed in i- or z-band.

For the four globular clusters analysed here, the gain brought by GeMS over the natural seeing is a narrower FWHM by a factor 1.6 to 2.8. 
This observed difference, although the observations were taken at a similar natural seeing, can be explained by different reasons (by increasing order of importance:
\begin{itemize}
\item  the NGS constellation is different :  the three NGS are used to compensate for the tip-tilt and tilt-anisoplanatism modes. Depending on their position over the field, and how they cover it, the correction will be more or less uniform.
\item  the laser photon return : if GeMS receives less laser photons, the loops are running at a slower frequency and the overall performance will decrease. This could be due to varying sodium density cite{Neichel2013}, airmass or laser power. 
\item  the turbulence profile : depending where the principal layers are located, GeMS can more or less correct them. 
\end{itemize}

As an illustrative example, we took a look in particular at NGC4590. This globular cluster was observed on 2012-03-15UT at an elevation of 73deg and an airmass of 1.043, and on 2012-05-10UT at an elevation of 75deg and an airmass of 1.035. The NGS constellation was the same for both observation. In this case, the factor 2 of difference in FWHM performance (see Table~\ref{tab:gc}) can come from either the turbulence profile either the laser return. It is known that in March there is statistically less sodium than in May. From Table 2 of \cite{Neichel2013}, the average sodium return in March 2012 was 6.5 photons/s/cm2/W versus 13 photons/s/cm2/W in May 2012. Looking at the $\mathrm{C}_{\mathrm{n}}^{2}$ profile for these two nights (see Figure~\ref{fig:Cn2}), the March night has 1.74  $\mathrm{C}_{\mathrm{n}}^{2}.\mathrm{dH}.10^{13} \mathrm{m}^{1/3}$ at 0km while the May night has 4.11 $\mathrm{C}_{\mathrm{n}}^{2}.\mathrm{dH}.10^{13} \mathrm{m}^{1/3}$ at the same altitude. There is also a factor 2 of $\mathrm{C}_{\mathrm{n}}^{2}$ turbulence at 1.69km with 0.72 $\mathrm{C}_{\mathrm{n}}^{2}.\mathrm{dH}.10^{13} \mathrm{m}^{1/3}$ for March 2012 versus 1.15 $\mathrm{C}_{\mathrm{n}}^{2}.\mathrm{dH}.10^{13} \mathrm{m}^{1/3}$ for May 2012.
It seems then that the most influential factors for the FWHM performance while observing these datasets were the level of ground layer turbulence and the photon return from the LGS.

\vspace{-1cm}
%
%
%
\vspace{0.3cm}
\subsubsection{Improvement FWHM - Seeing}

%

\begin{figure*}
\begin{center}
\begin{tabular}{c c}
\includegraphics[width=9.4cm]{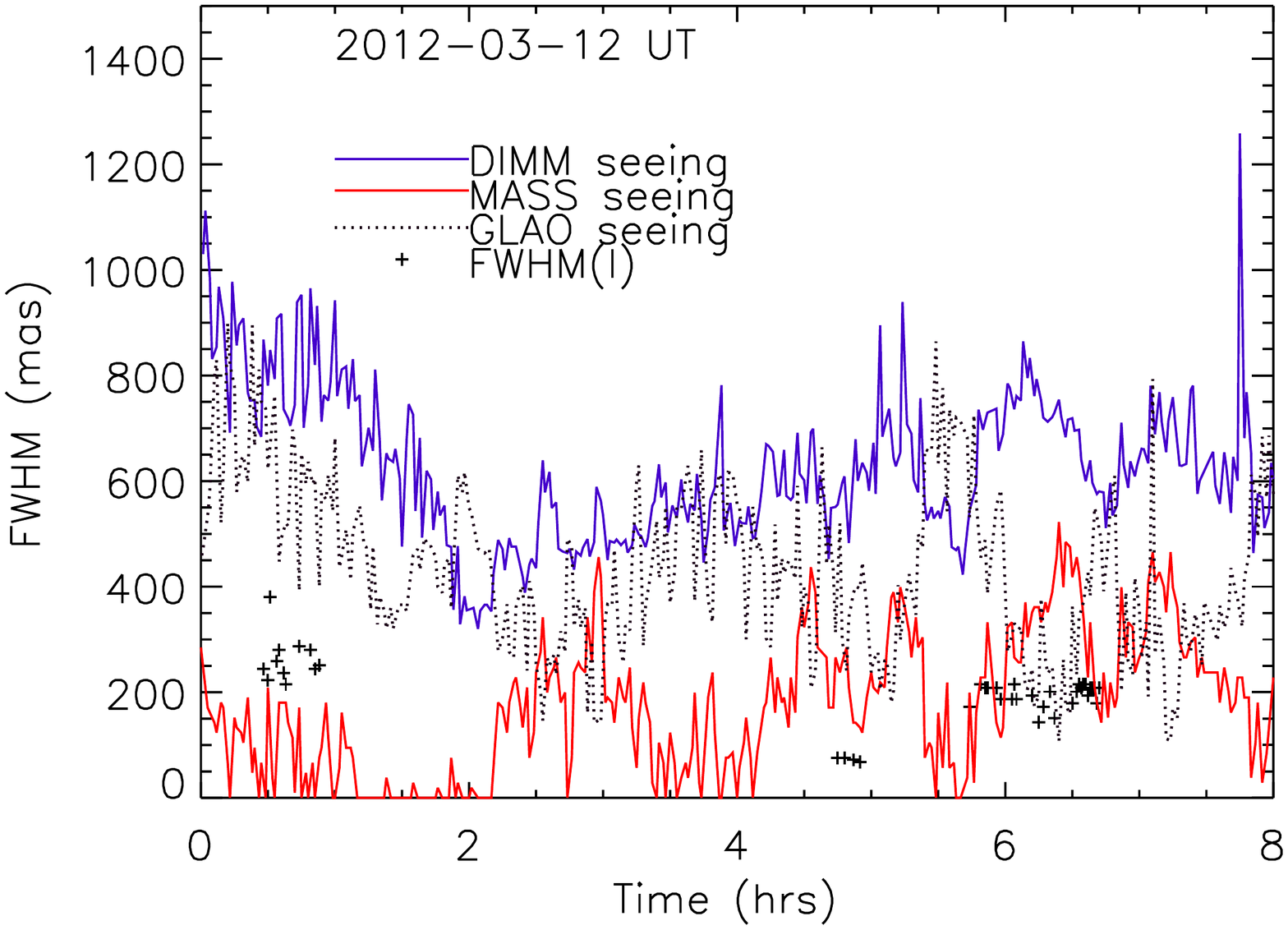} & \includegraphics[width=9.4cm]{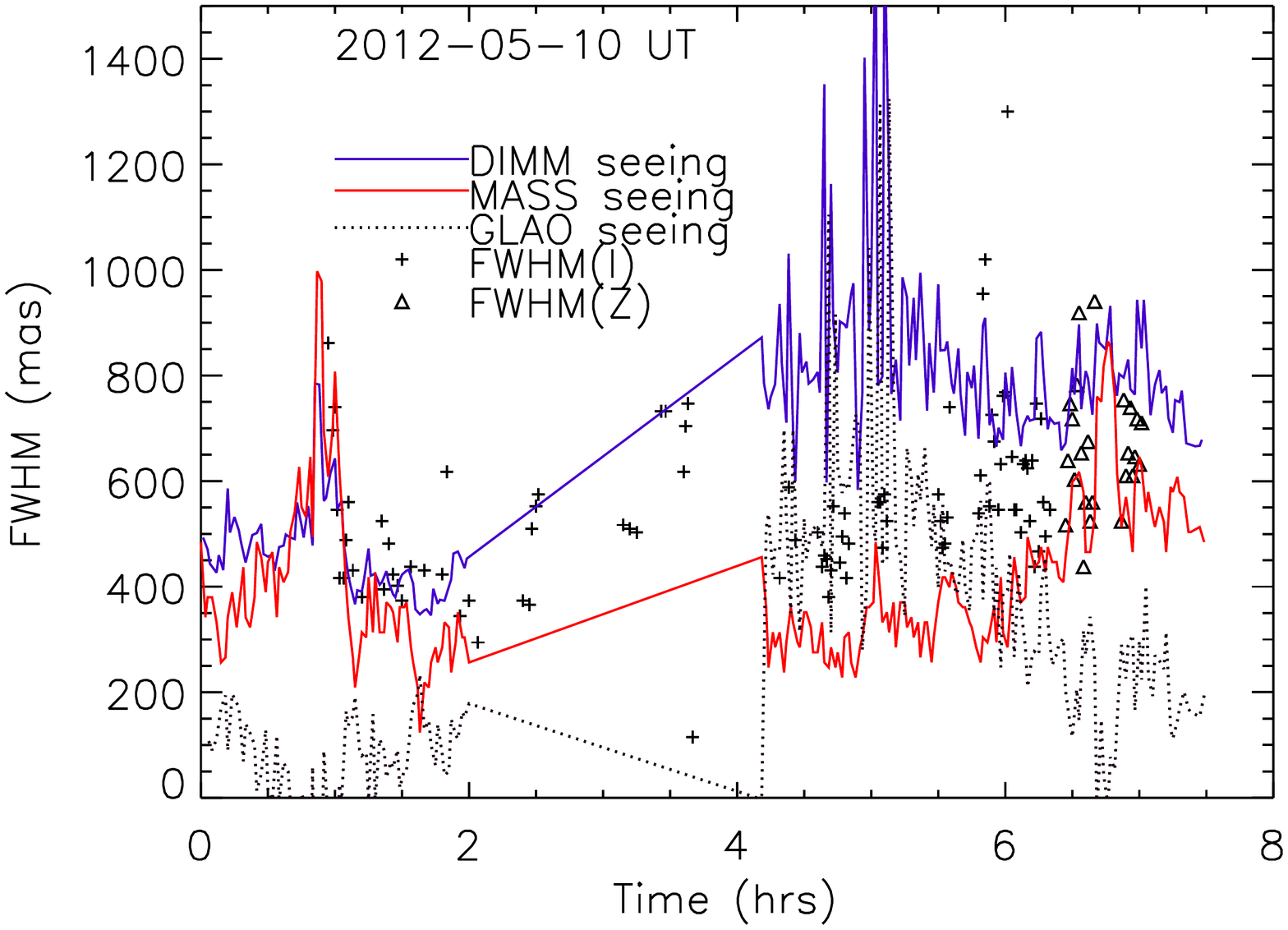}\\
\textit{a. 2012-03-12 UT } & \textit{b. 2012-05-10 UT}\\
\end{tabular}
\caption{Atmospheric conditions on two of the GeMS+GMOS nights. The blue lines show the total (DIMM) seeing, the red lines the free-atmosphere seeing (MASS) and the black dotted lines the GLAO seeing. The crosses are the FWHM of i-band GeMS+GMOS images, the diamonds the FWHM of z-band GeMS+GMOS images. The errors on the FWHM is $\pm$ 20 mas. Left : during this March 2012 night, we obtained better performance than a GLAO system will provide us. Right : during this May night, FWHM performance equivalent to the ones expected by a GLAO system were obtained between 4hrs and 6hrs, although the GeMS+GMOS FWHM performance were very variable. }
\label{fig:seeing}
\end{center}
\end{figure*}

The Differential Image Motion Monitor (DIMM) is located at 1.5m above the ground on Cerro Pach\'on and delivers an estimate of the seeing in the total atmosphere, normalized at 500nm. It is accompanied by a Multi-Aperture Scintillation Sensor (MASS) permiting one to measure the seeing in the free atmosphere above $\sim$0.5km. The ground layer seeing produced in the first 0.5km above the observatory can be evaluated by subtracting the turbulence integrals measured with the DIMM and MASS \citep{Tokovinin2007}. The FWHM PSF of a GLAO system is therefore expected to be close to the free atmosphere seeing but never better as it cannot correct for free atmosphere turbulence.\\
The average FWHM presented in Figure~\ref{fig:seeing} was obtained  from GeMS+GMOS images taken in i-band ($\lambda_{\mathrm{c}}=$780nm). To compare the FWHM to the seeings, we corrected the DIMM and the MASS seeing values from a factor of 0.915, following the Equation 5 from \cite{Tokovinin2002}. 

Figure~\ref{fig:seeing} shows the evolution of total and free-atmosphere seeings measured by the DIMM and MASS site monitor at Cerro Tololo on two nights. These measurements are a good approximation of the atmospheric conditions at Cerro Pach\'on as these two telescope sites are separated by 14km and have a 300m altitude difference. This figure also allows us to visualise the improvement factor between natural seeing, GLAO seeing and the average FWHM for these two different observing nights, as an example for the two observing periods with GeMS+GMOS. \\
We can see in Figure~\ref{fig:seeing} a., for the 2012-03-12 UT night, the FWHM values are below or equal to the GLAO seeing values. We are obtaining better if not equal performance than a GLAO system. The MCAO and the two Deformable Mirrors (DMs) have been correcting better than one DM could have done alone. 
However, Figure~\ref{fig:seeing} b., for the 2012-05-10 UT night, the FWHM results are mostly located between the DIMM and MASS data and during the third quarter of the night, these results are equivalent to the GLAO seeing values. We were able to obtain correction similar to the ones expected by a GLAO system. The main turbulence factor was then coming from another turbulence layer or GeMS performance were not optimal.

\subsubsection{NGS constellation comparison}
During the night of 2012-05-10 UT, NGC6496 was observed using different GeMS configurations such two different NGS constellations.

Depending on the constellation geometry and guide star magnitude, the expected performance will be different. Best constellation are the ones that cover most of the field, and the more distant the stars are, the lower the plate scale error will be. Generally, at the first order, we want to maximize the area of the triangle delimited by the three stars. However, this has to be mitigated by the noise propagation, i.e. the magnitudes of the stars.

\begin{figure*}
\begin{center}
\begin{tabular}{c }
\resizebox{1.5\columnwidth}{!}{ \includegraphics {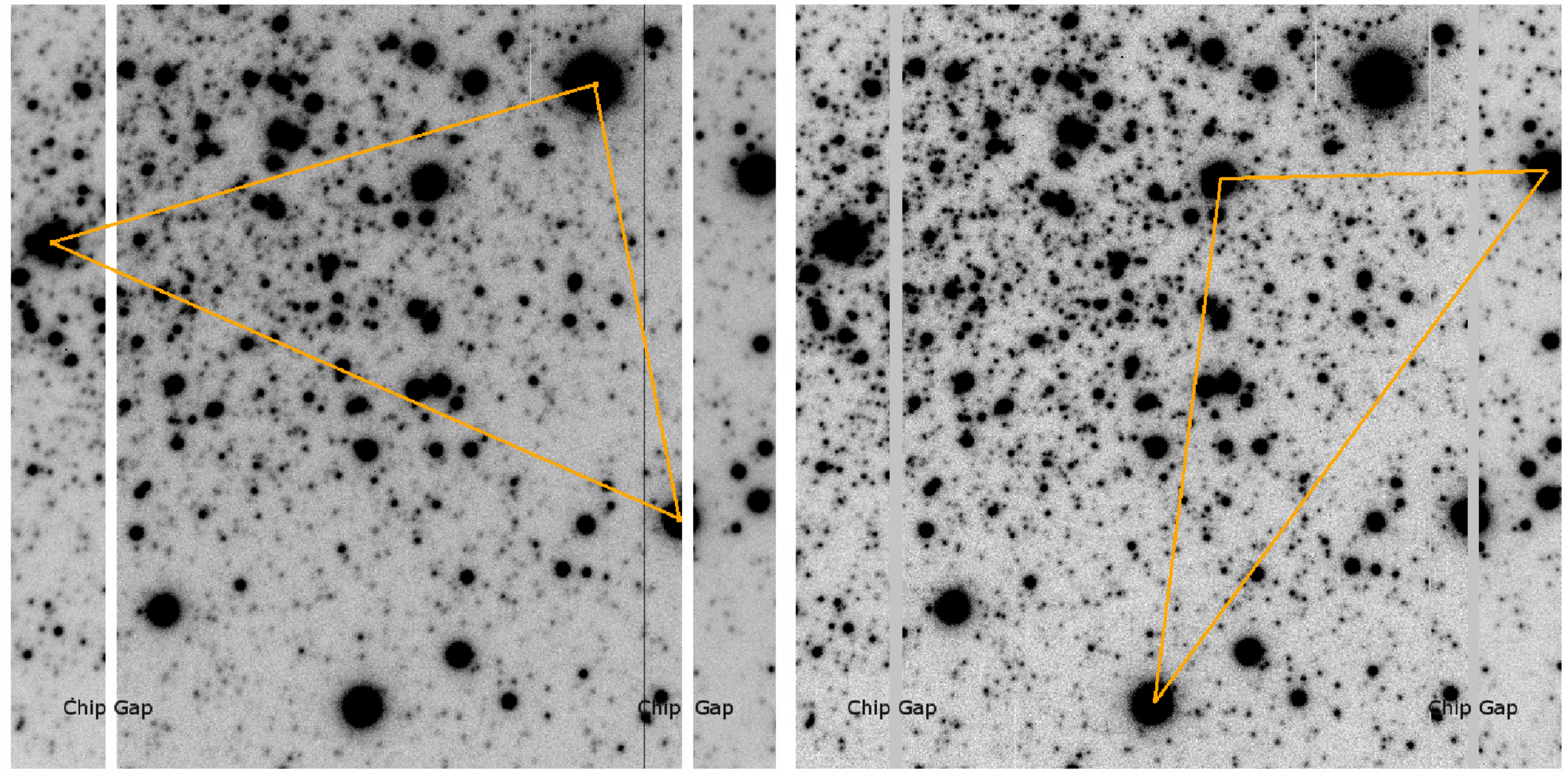}}\\
\resizebox{1.65\columnwidth}{!}{\includegraphics{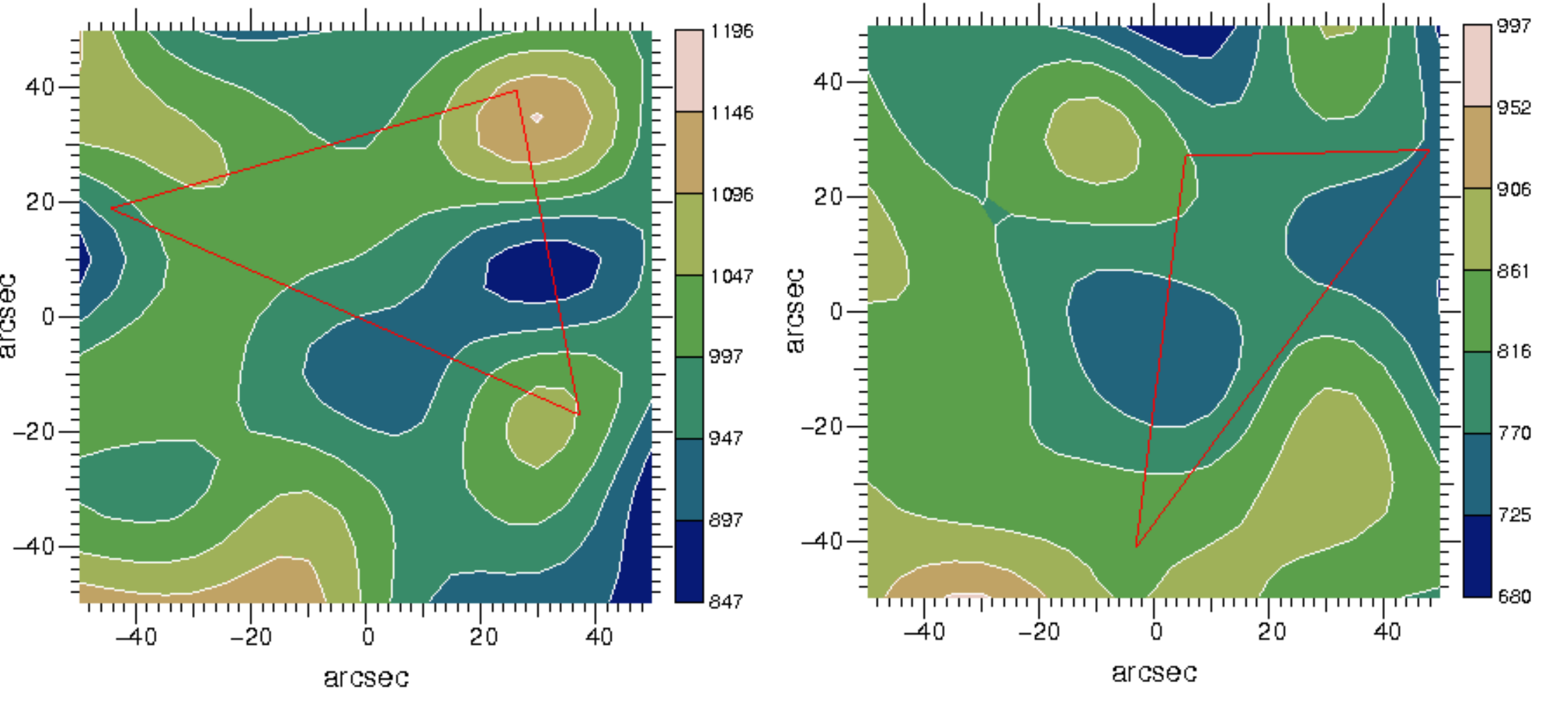}}\\
\end{tabular}
\caption{Comparison of FWHM performance obtained with different NGS constellation. Top : I-band 15sec individual observations of NGC 6496 using two different NGS constellations represented by the orange lines. Bottom : FWHM map corresponding of the above images. The FWHM unit is milli-arcseconds. }
\label{fig:comp}
\end{center}
\end{figure*}
One difference between both observations is the quality of the NGS constellation : looking at the top panel of Figure~\ref{fig:comp}  and at the r-mag given for the NGSs in Table~\ref{tab:comp}, it seems that the Top Left image has a NGS constellation which constitutes a better equilateral triangle with brighter NGSs and widely spread in the field. This implies bigger and more homogeneous FWHM zones as seen in the Bottom Left FWHM map (see Fig.~\ref{fig:comp}). \\
The NGS constellation for the Top Right image is limited to one part of the field and has fainter NGS r-magnitudes, which explains the more drastic separation of FWHM values in the Bottom Right map (see Fig.~\ref{fig:comp}).\\
Although the magnitude of the NGSs are different (see Table~\ref{tab:comp}), they are all in the range of NGS r-magnitude accepted for GeMS, i.e R$<$ 15mag. The Figure~\ref{fig:comp} shows that the GeMS system performed in equivalent way for both observations and our limiting factor appears to be the natural seeing. This is indeed not surprising as we already know that the GeMS system is under-dimensioned for the visible wavelength range, i.e it has an insufficient number of actuators and a too low loop frequency to be able to have an optimised correction with visible wavelengths. The GeMS+GMOS visible performance are therefore very dependent on the natural seeing and the atmosphere.\\
The GeMS LGS loop is currently limited by the combination of its laser power and format and is, in average, running at 300 to 400Hz. To compensate for the atmospheric turbulence distortions, the actual DM0 is using 240 actuators and DM9 is using 120 actuators. To perform as well in the red bands (r and i), any AO system would need at least twice the loop frequency (from 1 to 1.5kHz) and twice the current number of actuators. Moreover, a MCAO system based on LGS guide stars would require about 9 to 10 times more LGS spots on the sky and as many corresponding WFS. The laser power required to sustain this type of operation is about 25 to 30 times more than what the actual GeMS Lockheed Martin Coherent Technologies (LMCT) laser can deliver.

\begin{table*}
\caption{Table summarising the observations, the NGS magnitudes and the observing conditions for one particular target : NGC6496.}
\label{tab:comp}       
\begin{center}
\begin{tabular}{lllllllllll}
\hline
Images & Dates & Filter & Exp. Time & Aver. &CWFS1 &   CWFS2 R- &  CWFS3  & Airmass & $r_{\mathrm{0}}$ & MASS \\
& & & & FWHM & R-mag &   R-mag &  R-mag & & & seeing\\
\hline
Image 1 & 2012-05-10 UT & i-band & 15sec & 860 $\pm$ 168 mas  & 12.7 & 11.1 & 13.3 & 1.05 & 10 & 0.36'' \\ 
Image 2 & 2012-05-10 UT & i-band & 15sec & 691 $\pm$ 133 mas & 14.8 & 12.7 & 13.9 & 1.1 & 12 & 0.2'' \\ 
\hline
\end{tabular}
\end{center}
\end{table*}

\section{Throughput and zero-points estimation}
\subsection{Method}
We expect the GeMS throughput to be lower than GMOS simply due to the added number of mirrors in the MCAO system, each of which absorbs some of the transmitted light. What we aim to estimate is by how much the throughput deteriorates due to GeMS. We determined the difference of flux transmission between GMOS and the GeMS+GMOS (G+G) system using the following equation (Eq.~\ref{eq:1}):
\begin{equation} 
\mathrm{(G+G)\ Throughput} = \frac {\mathrm{(G+G)\ Flux}} {\mathrm{GMOS\ Flux}} \times \frac{\mathrm{GMOS\ Exp.Time}} {\mathrm{(G+G)\ Exp.Time}}
\label{eq:1}
\end{equation}

To obtain an absolute value of this throughput difference, we only created maps for data that we have equivalent GMOS images in gain, filter and binning. Such data exist only in the i-band. 

\subsection{Results and Discussion}
\label{results}

\begin{figure*}
\begin{center}
\begin{tabular}{c c}
 \includegraphics[width=6cm]{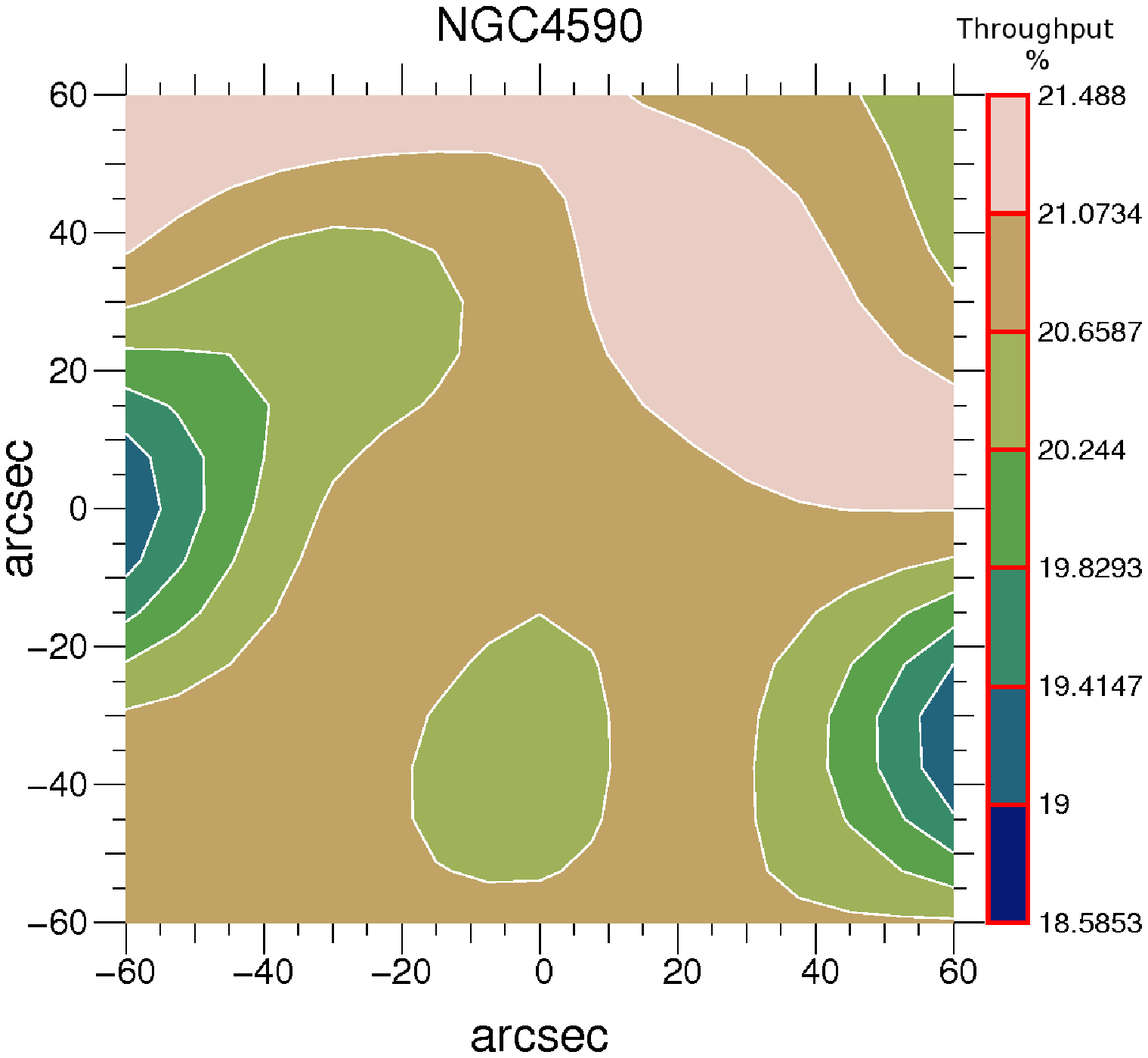} & \includegraphics[width=6cm]{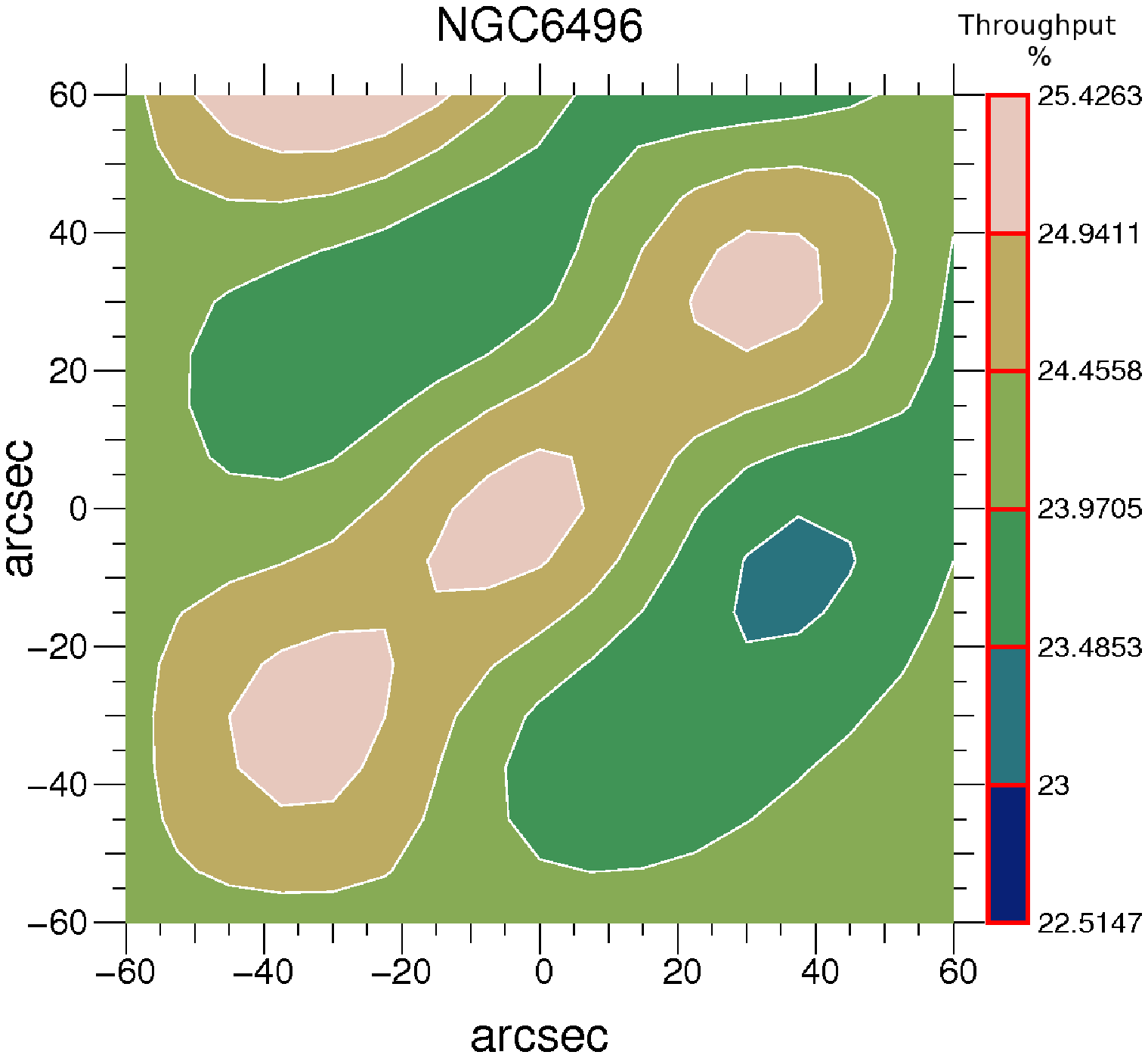}\\
\end{tabular}
\caption{Throughput difference map in percentage for two globular clusters : NGC4590 (Left) and NGC6496 (Right).}
\label{fig:phot}
\end{center}
\end{figure*}

Figure~\ref{fig:phot} shows the throughput difference in percentage for two globular clusters. Remarkably, for each individual frame, the GeMS throughput varies by less than 4.7\% across the examined portion of the CCD chips.

\begin{figure}
\resizebox{1\columnwidth}{!}{ \includegraphics{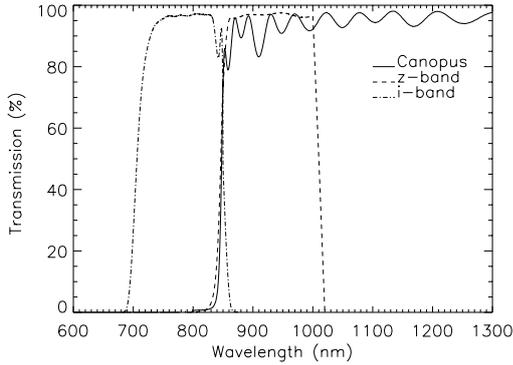} }
\caption{The solid line shows the Canopus transmission according to wavelength for the setup used in this work. It is evident that the beam-splitter cut-off point occurs within the i-band's 706-850nm range. This explains why the i-band throughput determined above is so much lower than we might expect from the GMOS throughput.}
\label{fig:canopus}       
\end{figure}

The low throughput observed in i-band can be explained by the wavelength cutoff of the Beam Splitter used in the AO bench. This beam splitter cuts at 850nm, all the light with a lower wavelength being sent into the Wave-Front sensors of the AO bench, only the wavelength larger than 850nm are directed toward the science instruments.
If GMOS would be used with GeMS for science operations, one option to improve the throughput would be to change the Beam Splitter (BS) with one having a cutoff at ~600nm (see Figure~\ref{fig:canopus}). This would allow us to observe in the GMOS r-band filter. We could not afford shorter wavelengths, as the laser sodium light (at 589nm) has to be seen by the AO Wave Front sensors. Note that changing the BS for a shorter wavelength cut will affect the limiting magnitude of the Natural Guide Stars, needed for the tip-tilt correction. We estimated that going for a 600nm BS would decrease the limiting magnitude by 1. This will also impact the sky coverage. Another possibility will be to use a beam splitter sending only the laser light ($\lambda$=589nm) to the wavefront sensor and the rest to GMOS. In that case, the tip-tilt sensing would be done with a peripheral WFS on the telescope guiding system. This solution would open observations with all the GMOS filters, however it would introduce more anisoplanatism in the images. Such a system is being implemented at Gemini North for the Altair AO system : 
the
current Altair science dichroic will be replaced by a sodium notch filter, passing
only the 589nm wavelength light from the LGS to the AO system \citep{Trujillo}. The
rest of the spectrum from 400 nm to the GMOS red cutoff at 1.1 microns
is intended as science capable light. Tip/tilt correction will be
performed close to the science target with the GMOS on-instrument
wavefront sensor or with the peripheral wavefront sensor. An image quality improvement of roughly a factor 2 is expected in this mode
over seeing limited observations.

During the May 2012 observing run, we also obtained GeMS+GMOS data for a photometric standard field \citep{Smith2007} in i- and z-band. Its exposure time is 12.5s at an airmass of 1.39.  The same field was observed in June 2012 with GMOS in the same filters with an exposure time of 4.5s at an airmass of 1.16.  Three stars, isolated from others field stars and from each other, were selected in this field and we measured their FWHM and flux. We used the \textit{FLUX$_{}$AUTO} from SExtractor. In both i- and z-bands, we gain in sensitivity in the GeMS+GMOS data, with an improvement factor 1.2 in i-band and 1.5 in z-band. In term of flux, as most of the i-band wavelength range is cut to feed the AO bench, we lose a factor of 2.9 on these three well isolated stars by using GeMS. However, in z-band, we gain a 1.5 factor by using GeMS on these same stars. The use of GeMS with GMOS is therefore useful not only for crowded fields but also for single well isolated stards.

\subsection{Zeropoint magnitude}
We also used the \cite{Smith2007} photometric fields observed with GeMS+GMOS to estimate the magnitude zeropoint in the observed bands : i-band, CaT-band and z-band, when observed. These fields were acquired in open-loop which causes a diminution of the light transmission. The photometric calibration was done on ten stars and we extrapolated the value for the CaT filter. \\
We obtained the following AB zeropoint magnitude values : $\textrm{mag}_{ZP}(\textrm{i-band}) = 29.00 \pm 0.15$ , $\textrm{mag}_{ZP}(\textrm{CaT-band}) = 28.85 \pm 0.20$, $\textrm{mag}_{ZP}(\textrm{z-band}) = 28.75 \pm 0.15$.

\section{Astrometric Performance}
Astrometry deals with the measurement of the positions of objects on the sky, with ultra high precision and as a
function of time. Based on the globular cluster data, we estimated a first astrometric performance of GeMS+GMOS.\\
\subsection{Distortion correction}
For a set of non-dithered images, it is expected that there will be no systematic distortions between images since the distortion map (if any) will remain in the same. As such, the presence of systematic distortions in images taken with GeMS would suggest the AO system introduced these distortions.
The distortion factors which may affect our dataset are:
\begin{itemize}
\item the (x,y) image offsets (in undithered images),
\item the image rotation, 
\item the astigmatism at 0$^\circ$ and 45$^\circ$, 
\item the focus,
\item and the higher order distortions.
\end{itemize}
The distortions were measured in the images by using first and second order Zernike polynomials.\\
To ensure the distortion correction algorithm performed adequately, a number of simulated images with various distortions were created and processed. The simulated images were created with a single type of distortion as well as a combination of all forms. The results from the simulated images showed the algorithm was sufficient, successfully removing all simulated distortion patterns.\\
For our undithered dataset on the target NGC3201, we have an average displacement of 23.49mas before distortion correction. After correcting for the first order distortions, we obtained an average displacement of 15.08mas.

\subsection{ Astrometry error}
\begin{figure*}
\begin{center}
\begin{tabular}{c c c}
 \includegraphics[width=5.5cm]{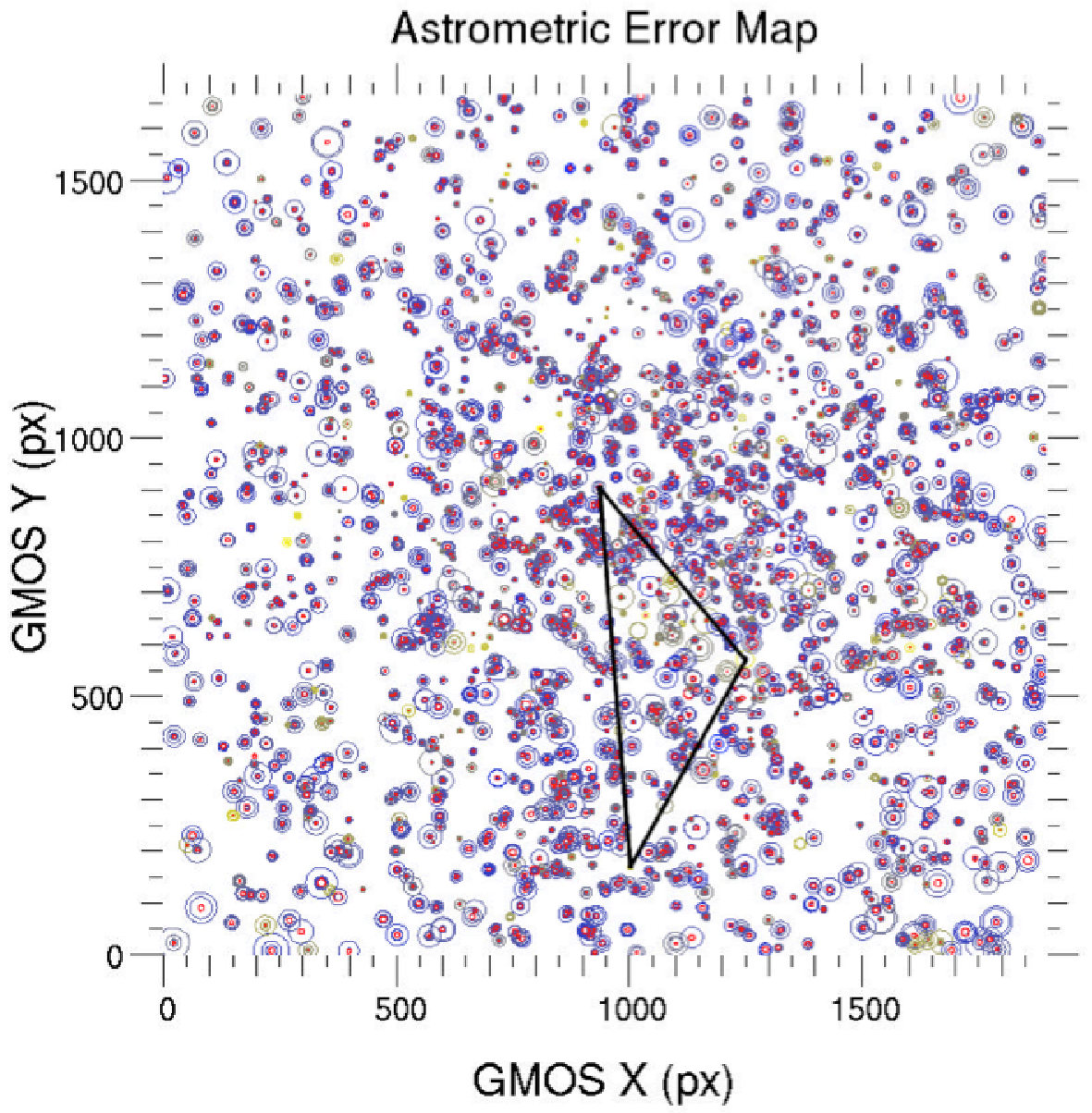} & \includegraphics[width=5.5cm]{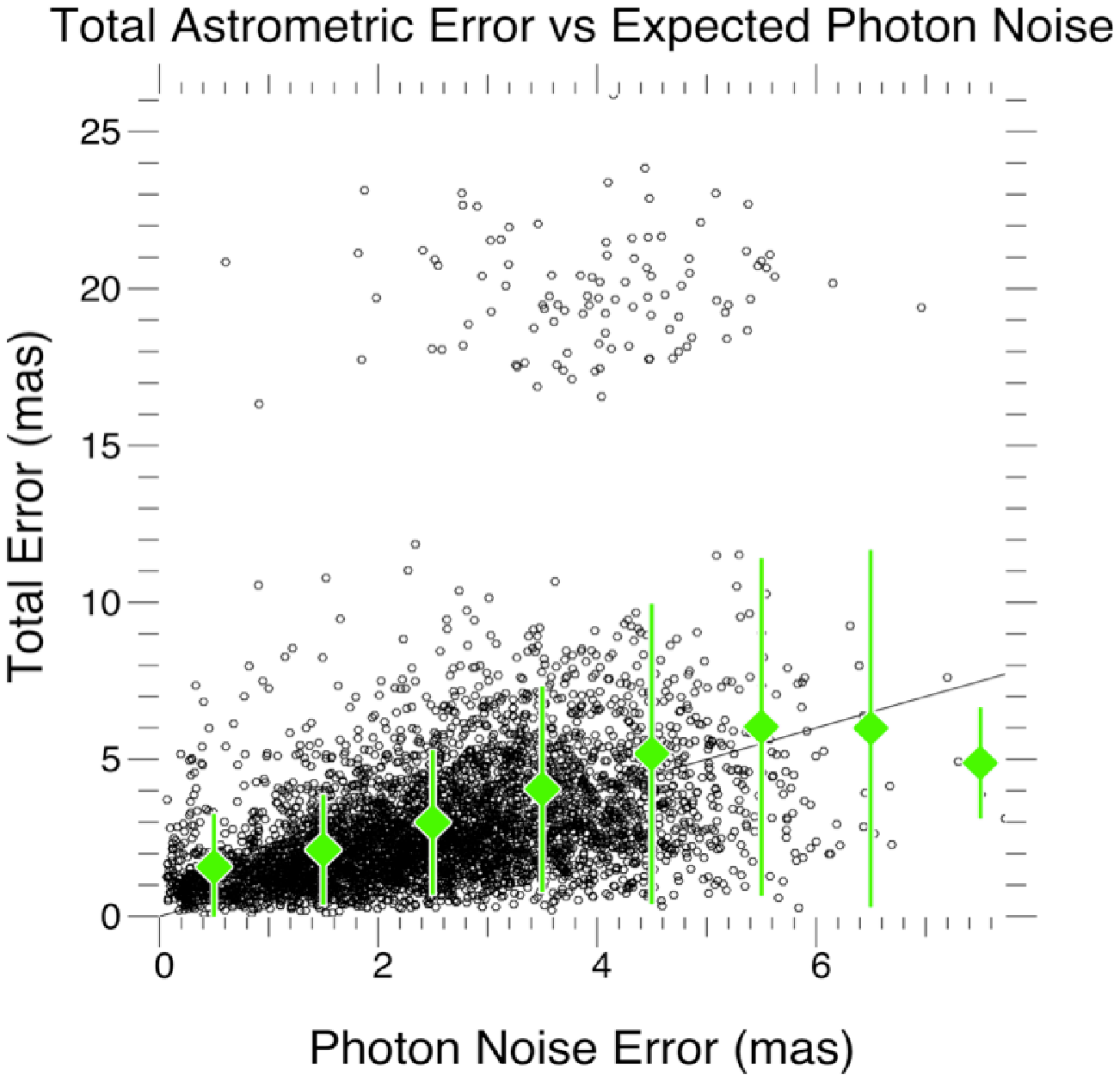} & \includegraphics[width=5.5cm]{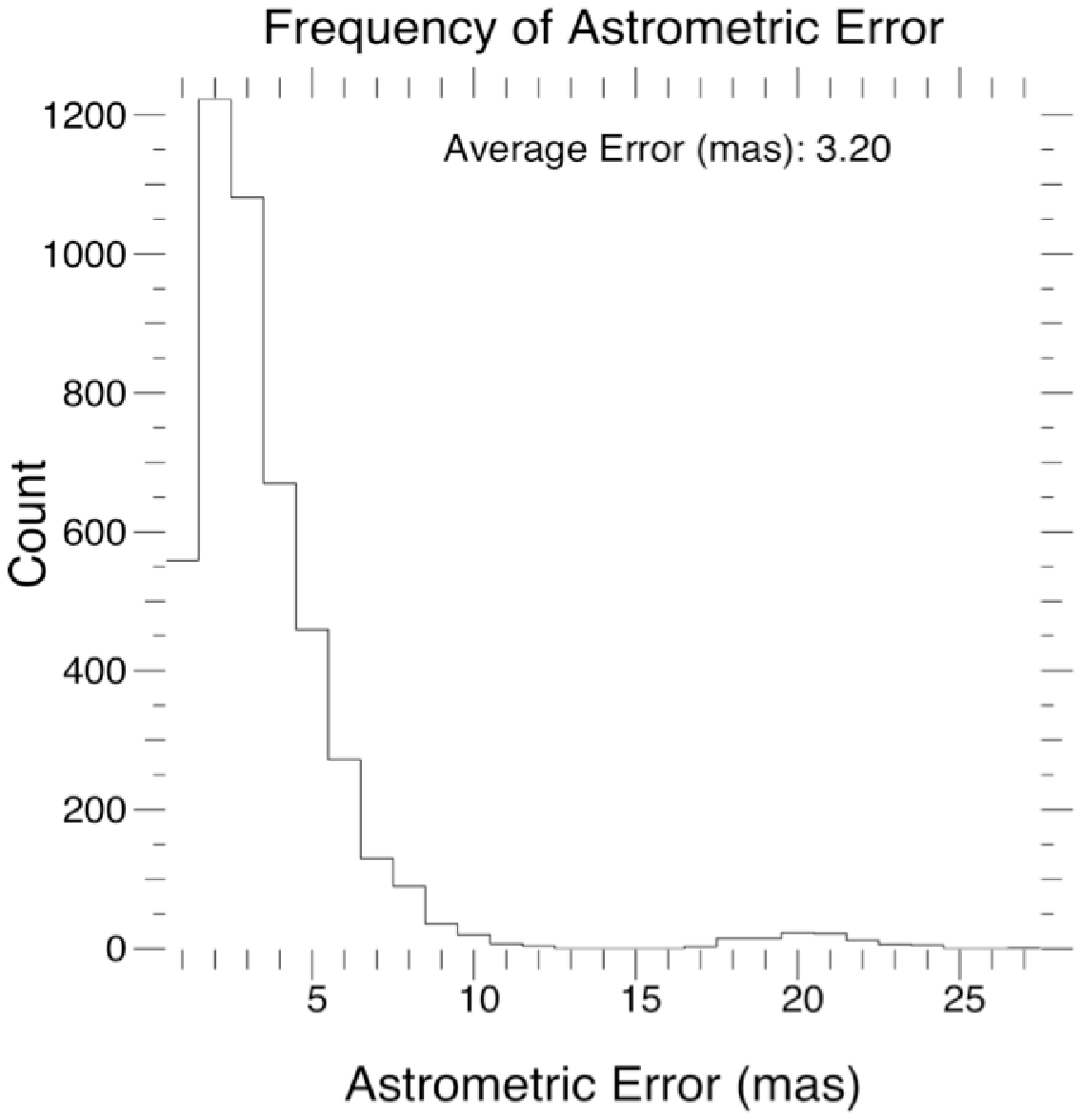}\\
\end{tabular}
\caption{Left : Absolute astrometric error map. Each circle is centered on a star. Yellow circles are a high flux stars, blue circles are the low flux star. The red circles represente the estimated photon noise. Center : Astrometric error versus the expected photon noise. Right : Histogram showing the frequency of the astrometric error.}
\label{fig:aa}
\end{center}
\end{figure*}

The average astrometric error for our different targets was determined only for the undithered datasets and are presented in Table~\ref{tab:aaerror}.\\
We are detailing here the case of NGC 4590 for which we have in our hands 31 individual images taken during the same night with the same configuration (exposure time, filter and binning). After finding the star position in each individual frame with SExtractor, we create a Master Reference Frame (MRF) from the average star position. We compared then the difference in position from all the individual images to the MRF. The results are shown in different ways: the comparison of the total astrometric error to the expected photon noise (Figure ~\ref{fig:aa} center), and the frequency of the astrometric error (Figure ~\ref{fig:aa} right). The photon noise is estimated following the Equation~\ref{eq:2}, which gives the error in the position of the center of the stars purely based on photon noise. 
For NGC4590, the average astrometric error is 3.20 mas.\\
The same method was applied to each globular cluster : a MRF was created for each of them. The number of undithered images is different for each target. 
Table~\ref{tab:aaerror}  summarizes the astrometric perfomance expected for undithered images. This performance varies between 2.4 and 7.4 mas, which is quite encouraging. We can also observe in Table~\ref{tab:aaerror} that there is no trend betwen the astrometric precision obtained and the numbers of undithered images used for the calculation . As seen from Figure ~\ref{fig:aa} (center), most of the error do follow the photon noise trend, hence no systematic errors seem to be present, at least for undithered data.
\begin{eqnarray} 
\sigma_{photon} \varpropto \frac {FWHM} {\sqrt{N_{photon}}}
\label{eq:2}
\end{eqnarray}

\begin{table*}
\caption{Table summarising the absolute astrometric error measured in one of the undithered dataset for our targets. \textbf{The number of undithered frames used for the astrometric calculation is also included in the table.}}
\label{tab:aaerror}       
\begin{center}
\begin{tabular}{|c|ccccccccccc|}
\hline
Targets & NGC & NGC & NGC & NGC & NGC & NGC &NGC  & CENTAU- & CIRCI- &M93 & SAGIT- \\
 & 2849 & 3201 & 3244 & 4590 & 5286 & 5408 & 6496 & RUS & NUS & & TARIUS\\
\hline
Abs. astro. & 2.47 & 4.73 & 2.80 & 3.20 & 4.4 & 7.38 & 5.39 & 5.51 & 5.66 & 5.15 & 5.28 \\
error (mas) & & & & & & & & & & & \\
\hline
Nber. undith. images  & 12 & 12 & 17 & 31 & 11 & 4 & 6 & 17 & 21 & 5 & 6 \\
\hline
\end{tabular}
\end{center}
\end{table*}



\section{Discussion - Conclusion}

\subsection{Expected Performance}
During 2014, a CCD upgrade has occurred for GMOS-S. The EEV CCDs have been removed and Hamamatsu CCDs have been installed and commissioned.
These new CCDs are more sensitive in the red part of the visible spectrum : in i-band, the CCD Quantum Efficiency (QE) improved from 65\% to 90\%, which corresponds to a 1.38 factor improvement. In z-band, the CCD QE improved from 30\% to 85\%, corresponding to a  2.83 factor improvement. With the arrival of the new CCDs, we also installed a Y-band filter with a central wavelength of $\lambda_{\mathrm{c}}=1010\mathrm{nm}$ and covering the wavelength range [970-1070].  This same range is covered by the filter Z of the instrument GSAOI used exclusively with GeMS. The better red sensitivity of the Hamamatsu CCDs and this new near-infrared filter in GMOS allows us to envisage the continuation of the use of the GeMS+GMOS system at least for imaging data. \\
We expect to upgrade in 2015 the GeMS natural guide WaveFront Sensor to a more sensitive version and more robust version called NGS2. The main benefit will be a large increase (50\% sky coverage up to 40 degree from galactic plane with 3 guide stars, 50\% sky coverage at to galactic pole with one guide star) in the GeMS sky coverage, as GeMS will be able to guide on stars as faint as r-mag=17. 
 In 2016, GeMS will also receive a new laser: the goal is primarily to increase laser operation robustness in order to integrate GeMS in the regular Gemini queue system. A possible additional benefit might be an increase in laser photon return leading to improved AO performance. Gemini North is currently upgrading its AO facility ALTAIR to support GMOS in the visible bands with a new dichroic, we are considering how to replicate this effort for GeMS.

\subsection{Science Cases with GeMS+GMOS} 
Two of the advantages of AO science in the visible  are the availability of better science detectors in the visible, with lower dark current and lower read noise than the ones used in infrared, and the fact that the visible sky is much darker than the K-band sky.
The combination of the visible instrument GMOS with the MCAO
system GeMS is a unique opportunity and could have then an important
scientific impact, as a pathfinder for future extremely large telescope instrumentation.
Due to the correction of the crowding noise, the first and obvious application of such a system is
open/globular clusters. We will be able to better resolve stellar populations and
obtain deeper magnitude limit from ground-based telescope. This will
help with cluster classifications \citep{Gerashchenko2013}, age \citep{West2004, Bridges2006}, metallicity \citep{Vanderbeke2014}, distance and
reddening determination \citep{Bonatto2013}.\\
Nebulae, and more specifically planetary nebulae \citep{Zijlstra2002, Villaver2003}, would also take
advantage of this observing system in order to characterise their weak
surrounding emissions and improve our understanding or their association
with star formation.\\
Galaxies and the study of their morphology \citep{Baillard2006, Kuminski2014, Dieleman2015}, their disk formation, their relation with the intergalactic medium (IGM), and the link between the evolution of the IGM and star formation, for
examples, would too \citep{Scannapieco2006, Oppenheimer2009, Wiersma2010}.\\
Moreover, faint targets, such as distant galaxies and gravitational arcs
 \citep{Ellis2001, Hu2002, Glassman2002, Messias2014} are also an important area to explore with visible MCAO. A large field
of view with a great AO correction, and therefore a great improvement
in resolution, is a great combination for studying
the distant Universe.

\subsection{Conclusion}
We have in our hands the first MCAO visible data. \\
The astrometric and photometric performance level reached is very encouraging to deepen the study and develop the science capability of such a system.\\
The FWHM performance varies from 60mas to 700mas depending on the seeing, the atmospheric conditions, and the AO performance. But overall, it is an improvement over the natural seeing by a factor 2 to 3. In terms of throughput, the AO correction allows to improve the sensitivity by a factor at least 1.5, the AB zeropoint magnitude found are $\textrm{mag}_{ZP}(\textrm{i-band}) = 29.00 \pm 0.15$ , $\textrm{mag}_{ZP}(\textrm{CaT-band}) = 28.85 \pm 0.20$, $\textrm{mag}_{ZP}(\textrm{z-band}) = 28.75 \pm 0.15$.
Finally, the astrometric performance, in terms of residual star jitter, gives an avg. error of 3.2mas, for typical exposures of 5 seconds, and it scales as expected with the photon noise, which means that no systematic error are detected, at least in the data studied in this paper.
Thanks to the availability of HST images for the GeMS+GMOS observed globular clusters, we are able to reach an astrometric calibration with a precision around 100 mas.
 A CCD upgrade for GMOS-South has increased its performance for the wavelength range [600-1050]nm. We are then expecting better performance of GeMS+GMOS. The GeMS+GMOS combination is also very interesting when used in Long-Slit, Multi-Object  and Integral Field Spectroscopy modes. The gain in spatial resolution will not only allow us to use smaller slit size but since the exposure time to reach a given signal-to-noise ratio scales
roughly as the square of the image quality, the use of such a system
represent a substantial efficiency improvement, comparing to GMOS without AO. Spectroscopic performance will be presented in a future paper. Thanks to GMOS versatility, we can envisage the use of GeMS with the IFU mode (5"x7"), which can be used in a similar fashion as with the Narrow Field Mode (7.5"x7.5") of the system VLT/MUSE- GALACSI \citep{Strobele2012}.

\section*{Acknowledgements}
Part of this work has been founded by the Gemini Observatory. The Gemini Observatory is operated by the As- sociation of Universities for Research in Astronomy, Inc., under a cooperative agreement with the NSF on behalf of the Gemini partnership: the National Science Foundation (United States), the Science and Technology Facilities Council (United Kingdom), the National Research Council (Canada), CONICYT (Chile), the Australian Research Council (Australia), Ministreio da Ciencia e Tecnologia (Brazil) and Ministerio de Ciencia, Tecnologia e Innovacion Productiva (Argentina). 
Part of this work has been funded by the French ANR programme WASABI – ANR-13-PDOC-0006-01.
Some of the data presented in this paper were obtained from the Mikulski Archive for Space Telescopes (MAST). STScI is operated by the Association of Universities for Research in Astronomy, Inc., under NASA contract NAS5-26555. Support for MAST for non-HST data is provided by the NASA Office of Space Science via grant NNX09AF08G and by other grants and contracts.
This research has made use of the SIMBAD database, op- erated at CDS, Strasbourg, France. 
The authors would like to thank the anonymous referee for constructive comments which helped us to improve the precision and clarity of the paper. The authors are also grateful to the GeMS team, Andres Guesalaga and Angela Cortes for providing the data and the fruitful discussions about the Cn2 profiles, Mischa Schirmer and Roberto Munoz for helpful discussions.
\bibliographystyle{mnras}
\bibliography{draft13_mnras}

\bsp	
\label{lastpage}
\end{document}